\documentclass[prd,twocolumn,tightenlines,showpacs,nofootinbib,superscriptaddress,nobalancelastpage]{revtex4-1} 

\usepackage{subfigure}
\usepackage{graphics}     
\usepackage{graphicx}     
\usepackage{longtable}     
\usepackage{url}          
\usepackage{bm}           
\usepackage{natbib}
\usepackage{textpos}
\usepackage{amsfonts,amsmath,amssymb,mathrsfs,bm,amsthm}
\usepackage{float}
\usepackage{booktabs}
\usepackage{array}
\usepackage{tabu}
\usepackage{dcolumn}
\usepackage{rotating}
\usepackage{ulem}
\usepackage{soul}

\newcommand{\RN}[1]{%
  \textup{\uppercase\expandafter{\romannumeral#1}}%
}

\usepackage{nicefrac}
\usepackage{amsthm}
\usepackage{color}
\usepackage{cancel}
\usepackage{appendix}
\usepackage{makecell}
\usepackage{xcolor}
\usepackage[colorlinks=true,linkcolor=black,citecolor=blue,urlcolor=blue,bookmarksopen]{hyperref}

\newcommand{\appropto}{\mathrel{\vcenter{
  \offinterlineskip\halign{\hfil$##$\cr
    \propto\cr\noalign{\kern2pt}\sim\cr\noalign{\kern-2pt}}}}}
\renewcommand{\v}[1]{\boldsymbol{#1}}		

\begin{document}
\title{Pseudovector and pseudoscalar spin-dependent interactions in atoms}

\date{\today}

\author{Pavel Fadeev}
\thanks{pavelfadeev1@gmail.com}
\affiliation{Helmholtz Institute Mainz, Johannes Gutenberg University, 55128 Mainz, Germany}

\author{Filip Ficek}
\affiliation{Institute of Physics, Jagiellonian University, \L{}ojasiewicza 11, 30-348 Krak\'ow, Poland}

\author{Mikhail G.~Kozlov}
\affiliation{Petersburg Nuclear Physics Institute of NRC ``Kurchatov Institute'', Gatchina 188300, Russia}
\affiliation{St.Petersburg Electrotechnical University LETI,Prof. Popov Str. 5, 197376 St.Petersburg, Russia}

\author{Dmitry Budker}
\affiliation{Helmholtz Institute Mainz, Johannes Gutenberg University, 55128 Mainz, Germany}
\affiliation{Department of Physics, University of California at Berkeley, Berkeley, California 94720-7300, USA}

\author{Victor V.~Flambaum}
\affiliation{Helmholtz Institute Mainz, Johannes Gutenberg University, 55128 Mainz, Germany}
\affiliation{School of Physics, University of New South Wales, Sydney, New South Wales 2052, Australia}

\begin{abstract}
Hitherto unknown elementary particles can be searched for with atomic spectroscopy.
We conduct such a search using a potential that results from the longitudinal polarization of a pseudovector particle. 
We show that such a potential, inversely proportional to the boson's mass squared, $V \propto 1/M^2$, can stay finite at $M \to 0$ if the theory is renormalizable.
We also look for a pseudoscalar boson, which induces a contact spin-dependent potential that does not contribute to new forces searched for in experiments with macroscopic objects, but may be seen in atomic spectroscopy. We extract limits on the interaction constants of these potentials from the experimental spectra of antiprotonic helium, muonium, positronium, helium, and hydrogen.
\end{abstract}


\maketitle


\section{Introduction}
 
A possible explanation for various outstanding puzzles in physics, such as the origins of dark matter \cite{Bertone05} and dark-energy \cite{Friedland03,Flambaum09}, the strong-CP puzzle \cite{Moody}, and the hierarchy puzzle \cite{Graham05} is the existence of beyond-the-standard-model (exotic) bosons. The exchange of such virtual bosons gives rise to an interaction potential. This motivates experimental searches for such potentials in nuclear, atomic, and molecular phenomena \cite{Safronova,Stadnik2018,Jiao2019}.

Recent work \cite{Fadeev2019} derived a list of these potentials, sorted by types of interactions (as opposed to \cite{Dob}, which classified the potentials by their spin-momentum structure). These are nonrelativistic
potentials in coordinate space, induced by the exchange of spin-zero or spin-one exotic bosons between fermions. Reference \cite{Fadeev2019} lists two types of potentials that were omitted in \cite{Dob}:

(a) A potential proportional to the inverse square of the intermediate spin-one boson mass, originating from its longitudinal polarization.

(b) Potentials that include the contact term $\delta(r)$, with $r$ being the distance between the interacting fermions.

Point (a) is important for the study of exotic bosons with pseudovector-pseudovector interactions. Point (b) is of concern when an experimental search for new bosons is conducted in atomic systems, where a contact interaction can play a vital role. Next, we discuss each of these potentials and the methodology of using them to obtain constraints on the properties of new bosons. Then, in section III we use these potentials to obtain novel limits on boson mass and coupling strength in various atomic systems. We conclude in Section IV.
\\
\section{Properties of pseudovector and pseudoscalar potentials}
\subsection{Potential proportional to 1/$\mathbf{\mathit{M}^2}$}
Among the nine potentials derived in \cite{Fadeev2019} which describe the exchange of an exotic boson between two fermions or macroscopic objects, the pseudovector-pseudovector potential is the only  velocity-independent one with a term inversely proportional to the boson mass squared:
\begin{widetext}
\begin{align}
\label{pseudovector-pseudovector_potential}
&V_{AA}(\v{r}) =\\  &- g_1^A g_2^A \underbrace{ \v{\sigma}_1 \cdot \v{\sigma }_2 \frac{e^{-M r}}{4 \pi r} }_{\mathcal{V}_2} 
-  \frac{g_1^A g_2^A m_1 m_2}{ M^2} \underbrace{ \left[ \v{\sigma}_1 \cdot \v{\sigma }_2 \left[ \frac{1}{r^3} + \frac{M}{r^2} + \frac{4 \pi}{3} \delta(\v{r}) \right]  -  \left( \v{\sigma}_1 \cdot \hat{\v{r}} \right) \left( \v{\sigma }_2 \cdot \hat{\v{r}} \right)  \left[ \frac{3}{r^3} + \frac{3M}{r^2} + \frac{M^2}{r} \right]  \right] \frac{e^{-M r}}{4 \pi m_1 m_2} }_{\mathcal{V}_3} \, .\nonumber
\end{align}
\end{widetext}
Here, $g^A$ are dimensionless
interaction constants that parametrize the pseudovector  interaction strength, $\v{\sigma}_1$ and $\v{\sigma}_2$ denote the Pauli spin-matrix vectors of the two fermions, $m_1$ and $m_2$ are the masses of the fermions, $M$ is the mass of the boson, $\hat{\v{r}}$ is the unit vector directed from fermion 2 to fermion 1, and $r$ is the distance between the two fermions. We work in natural relativistic units, $\hbar = c = 1$. Parts of the potentials defined as $\mathcal{V}_2$ and $\mathcal{V}_3$ link these terms to the definitions of the potentials described in \cite{Dob}. While deriving $V_{AA} (\bold{r})$ we have retained the leading order spin-dependent terms; that is why operators such as $\mathcal{V}_8$ in \cite{Dob} do not show up in Eq. \eqref{pseudovector-pseudovector_potential}.

To find the interaction for composite systems, one should sum the interaction \eqref{pseudovector-pseudovector_potential} over all fermion constituents (electrons, protons, and neutrons), each with its own interaction constants. The result will be proportional to the nuclear or atomic spins,
similar to the usual magnetic interaction between atoms in a crystal. 
Examples of composite systems used in experimental searches for spin-dependent potentials can be found in Refs.\,\cite{Safronova,Leslie2014,Ji2017,Hunter2014,Kim2019}. 

\begin{figure} 
\begin{center}
\subfigure[]{\label{fig:bound2}\includegraphics[width=0.45\textwidth]{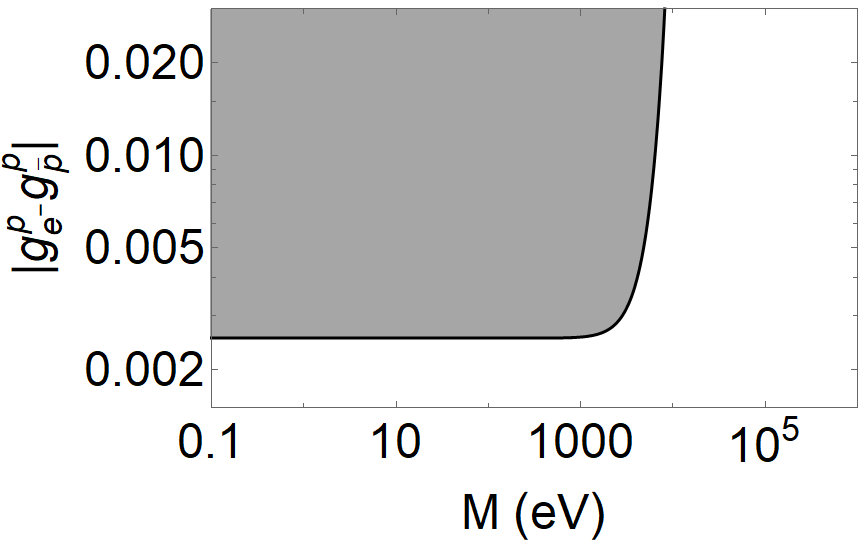}}
\subfigure[]{\label{fig:bound3} \includegraphics[width=0.45\textwidth]{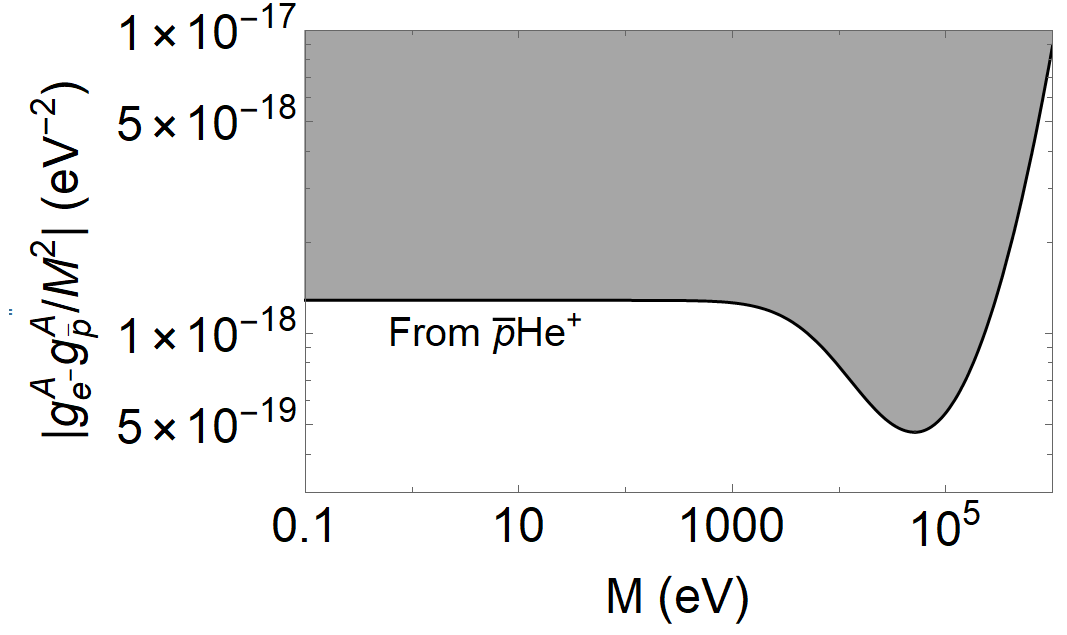}}
\end{center}
\caption{Constraints for the interaction between an electron and an antiproton at 90\% confidence level on the coupling constants as a function of boson mass. We are using states in the $(n,l) = (37, 35)$ manifold of antiprotonic helium $\bar{p}$He$^+$. The plots are based on the experimental data from \cite{Pask2009}, theoretical calculations from  \cite{Korobov2001}, and our numerical estimate of the spin-dependent contribution.
(a) Using the $V_{pp}$ potential of Eq.~(\ref{pseudoscalar-pseudoscalar_potential_form2}) in numerical integration. For $M < 10^2$\,eV the limit is at $0.0025$. 
(b) Using $V_{AA}$ in Eq.\,\eqref{pseudovector-pseudovector_potential}.
In the range $M < 10^{2}$ eV the bound is $g_e^A g_{\bar{p}}^A / M^2 \leq 1.3 \times 10^{-18} \,\text{eV}^{-2}$.
In the vicinity of $M=5 \times 10^4$ eV the bound is at $4.7 \times 10^{-19}$ eV$^{-2}$. This and other bounds are summarized in Table \ref{Table:summary}.} 
\end{figure}

\begin{figure} 
\subfigure[]{\includegraphics[width=0.46\textwidth]{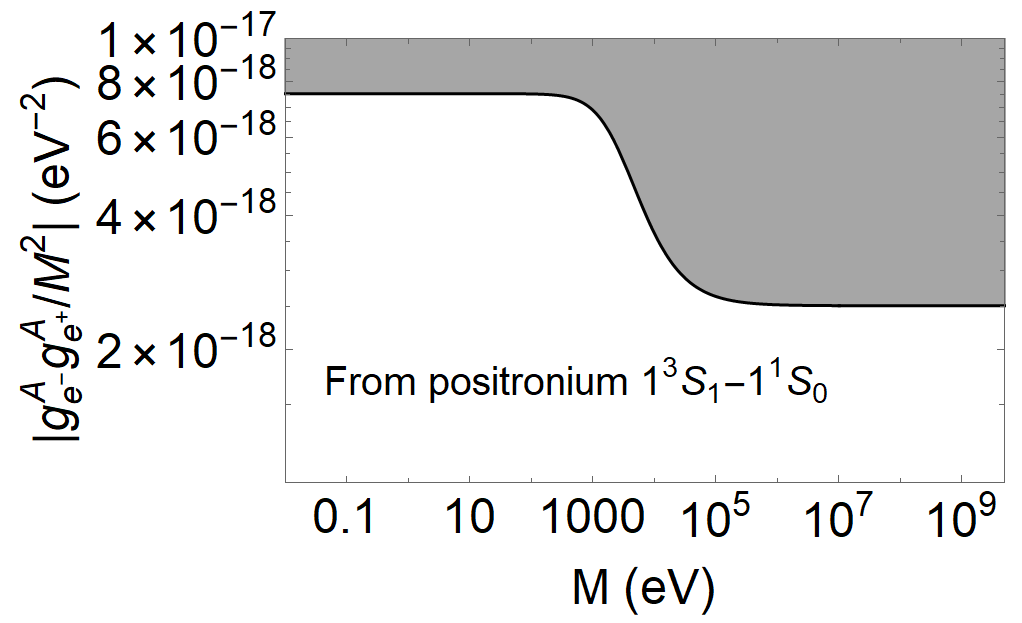} \label{fig:Positronium}}
\subfigure[]{\includegraphics[width=0.45\textwidth]{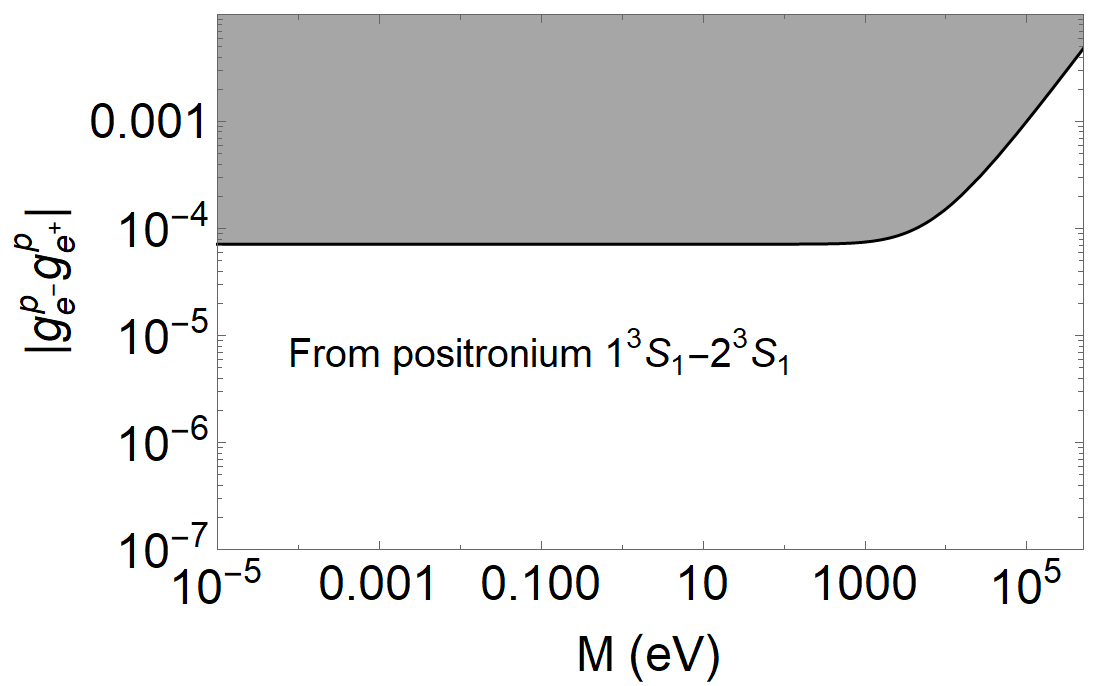}
\label{1s2sa}}
\subfigure[]{\includegraphics[width=0.45\textwidth]{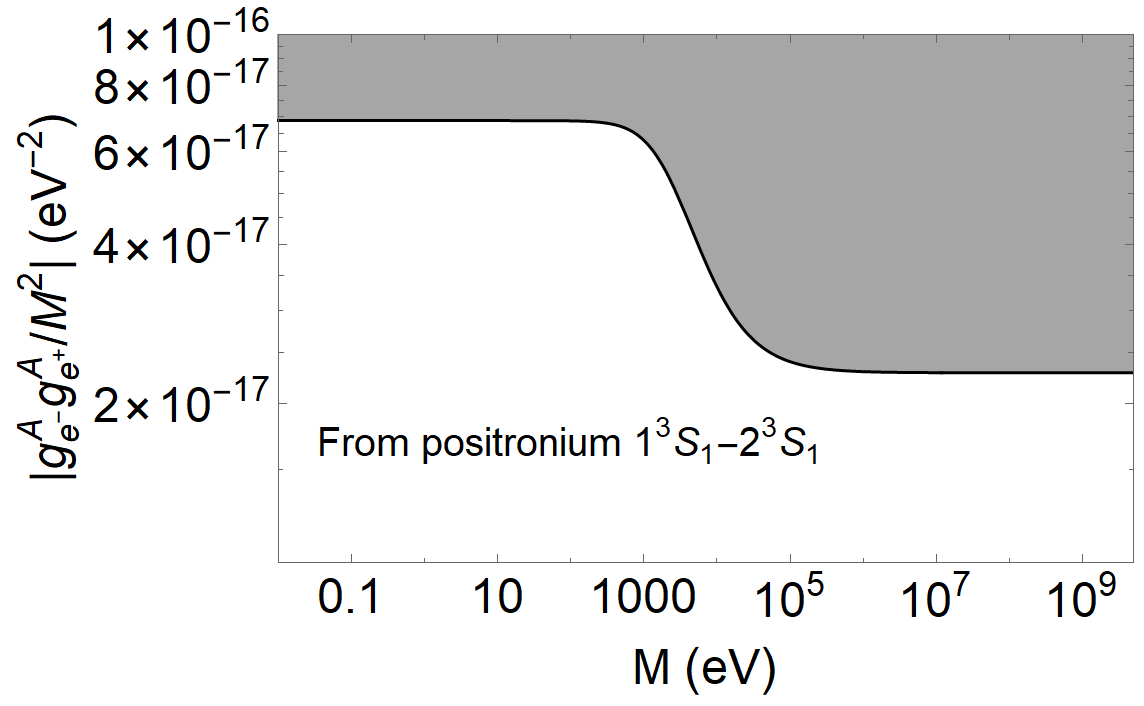}
\label{1s2sb}}
\caption{Constraints for the electron-positron interaction at 90\% confidence level on the coupling constants as a function of boson mass.
(a) The plot is based on experimental \cite{Ishida2014,Mills1975,Ritter1984} and theoretical \cite{Eides2014} values for the $1 ^3S_1-1 ^1S_0$ ground state transition in positronium \cite{Derek2015} and our numerical estimate of the spin-dependent contribution. 
(b)
The plot is based on experimental \cite{Fee1993} and theoretical \cite{Czarnecki1999,Manohar2000} values for the $1^3 S_1 - 2^3 S_1$ transition in positronium \cite{Fru19} and our numerical estimates of the spin-dependent contribution. The bound is based on
$V_{pp}$ potential of Eq.~\eqref{pseudoscalar-pseudoscalar_potential_form2}. 
(c)
Using $V_{AA}$ in Eq.~\eqref{pseudovector-pseudovector_potential}. Same transition as in (b).
} 
\end{figure}

The $\mathcal{V}_3$ term in Eq.\,\eqref{pseudovector-pseudovector_potential} arises from a longitudinal polarization mode for a massive spin-1 boson (which gives the term $q_{\nu}q_{\mu}/M^2$ in the massive vector boson propagator, $q_{\nu}$ being the four-momentum transferred between the fermions) 
and nonconservation of the axial-vector current ($q_{\nu} J^a_{\nu} \neq 0$) \cite{Fadeev2019,Malta,Parikh2021}. 
This term appears to have a singularity in the limit of the boson mass $M \to 0$. 
However, there should be no divergence in a renormalizable theory. 
Let us reflect on the following scenario based on the standard-model Lagrangian. 
We will see that as $M \to 0$, the combination of parameters $g_1^A g_2^A / M^2$ remains finite.
Consider $Z$-boson exchange between two fermions, where, in this case, the $Z$ boson has purely pseudovector interactions and does not mix with the photon [$\sin(\theta_W) = 0$, where $\theta_W$ is the weak mixing angle]. 
Then, the $Z$-boson mass is given by $M = gv/2$, where $v$ is the Higgs vacuum expectation value and $g$ is the (universal) electroweak interaction constant \cite{Gordon17}. The ratio $g^2/M^2 = 4/v^2$ remains finite as $M \to 0$, since the right-hand side is a constant.
For $v$ to be nonzero the fermion mass $m_f = fv/\sqrt{2}$ ($f$ is a species-dependent interaction constant) should be nonzero.
Thus, it is appropriate to place constraints on $g_1^A g_2^A / M^2$ of the $\mathcal{V}_3$ term in Eq.\,\eqref{pseudovector-pseudovector_potential}.
The association with renormalizability (with the Higgs mechanisms of mass generation) makes this case worthy of  experimental study. 
\\
\indent In the special case of a massless vector boson, $M = 0$,  only the $\mathcal{V}_2$ term remains in Eq.\,\eqref{pseudovector-pseudovector_potential} because a massless vector boson does not have a longitudinal polarization mode, and so the $\mathcal{V}_3$ term does not appear in this case.
\\ \\
\subsection{Bounds on contact terms}
Searches for exotic spin-dependent forces have been conducted both in atomic-scale experiments and in macroscopic-scale experiments \cite{Smorra19,Heckel06,Heckel08,Hunter13,Leslie2014,Kim2018,Ji2017}.
To search for new bosons, one may look for the difference between observations and theoretical predictions in the spectrum of an atomic, molecular, or nuclear system \cite{Ficek2017,Ficek2018, Frugiuele21}. Such difference can be due to an exotic-boson exchange between the system's constituents.

Unlike in macroscopic searches for new bosons, a contact term in a potential is of significance in atomic systems. Let us focus on determining a bound on the properties of spin-zero or spin-one exotic bosons by using a potential that includes the contact term $\delta(\v{r})$, such as the one appearing in Eq.\,\eqref{pseudovector-pseudovector_potential} and other potentials in \cite{Fadeev2019}. Contact terms were omitted in Ref.\,\cite{Dob}, but appeared in Refs.~\cite{Moody,Malta}.

As in \cite{Ficek2018}, we compare experimental results for the hyperfine structure of the antiprotonic helium \cite{Pask2009} with theoretical  QED-based calculations for this system \cite{Korobov2001}.
The difference between experiment and theory  $\Delta E$ at 90\% confidence level determined from
\begin{align} \label{deltaE90}
    \int_{-\Delta E}^{+\Delta E} \frac{1}{\sqrt{2 \pi}\sigma} e^{-(x-\mu)^2/(2\sigma^2) dx} = 0.9\, ,
\end{align}
where $\mu$ is the mean difference between theoretical and
experimental transition energies and $\sigma$ is the total uncertainty, $\sigma^2=\sigma_{th}^2+\sigma_{exp}^2$. 
To avoid misunderstanding, note that here theory uncertainty means uncertainty in the results of the  calculations of the transition frequencies within the standard model. 

We focus on a transition with the antiproton in the (n, l) = (37, 35) state and the electron in the (1, 0) state (where the first number is the principal quantum number, and the second one is the orbital angular momentum).
Let us consider the pseudoscalar-pseudoscalar potential, which appears in \cite{Ficek2018} and contains a contact term:
\begin{widetext}
\begin{equation}
\label{pseudoscalar-pseudoscalar_potential}
V_{pp}(\v{r}) = -  \frac{g_1^p g_2^p}{4} \underbrace{ \left[  \v{\sigma}_1 \cdot \v{\sigma }_2 \left[ \frac{1}{r^3} + \frac{M}{r^2} + \frac{4 \pi}{3} \delta(\v{r}) \right]
- \left( \v{\sigma}_1 \cdot \hat{\v{r}} \right) \left( \v{\sigma }_2 \cdot \hat{\v{r}} \right)  \left[ \frac{3}{r^3} + \frac{3M}{r^2} + \frac{M^2}{r} \right]   \right] \frac{e^{-M r}}{4 \pi m_1 m_2} }_{\mathcal{V}_3} \, .
\end{equation}
\end{widetext}

We deduce the contribution of this potential to the transition energies of the antiproton in antiprotonic helium.
The difference between the expectation values of $V_{pp}$ in the two states 
gives an estimate of the energy shift between the states caused by $V_{pp}$. 
The contact term contribution
is of the form $g_1^p g_2^p C$ where $C$ is a constant. Other terms in the expectation value of $V_{pp}$ vary with boson mass. We denote such terms by $g_1^p g_2^p \Delta U (M)$. Assuming the difference between theory and experiment $\Delta E$ at 90\% confidence level [Eq. \eqref{deltaE90}] is due to $V_{pp}$, we may write
\begin{equation} \label{smallerthanE}
    | \, g_1^p g_2^p \left( C + \Delta U (M)\right) \, | \leq  | \, \Delta E \, | \, ,
\end{equation}
which results in
\begin{equation} \label{plotcreation}
    | \, g_1^p g_2^p  \, | \leq \Big|  \frac{ \Delta E}{\left( C + \Delta U (M)\right)} \Big|  \, .
\end{equation} 
The left-hand side in this expression is the ordinate in Fig. \ref{fig:bound2}. In the regime  $C \gg \Delta U(M)$ the right-hand side would be a constant independent of $M$.
However, in the limit of large $M$ we obtain $\Delta U(M) \to  -C$ and nearly cancel it.
This may lead to a numerical instability at large $M$, discussed in Appendix \ref{ExclusionPlot}.

 \begin{figure} 
\includegraphics[width=0.46\textwidth]{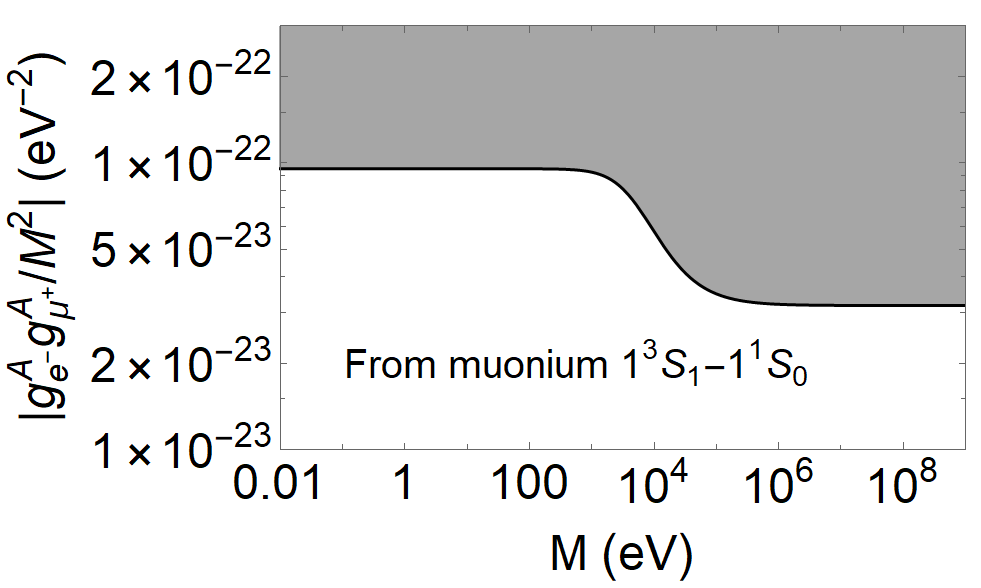}
\caption{Constraints for the interaction between an antimuon and an electron, at 90\% confidence level on the coupling constants as a function of boson mass, using $V_{AA}$ in Eq.~\eqref{pseudovector-pseudovector_potential}.
The plot is based on experimental \cite{Liu1999} and theoretical \cite{Karshenboim2014,Karshenboim2005} values for the hyperfine ground state transition in muonium \cite{Fru19} and our numerical estimate of the spin-dependent contribution. 
} 
\label{fig:Muonium}
\end{figure}

 \begin{figure} 
\includegraphics[width=0.46\textwidth]{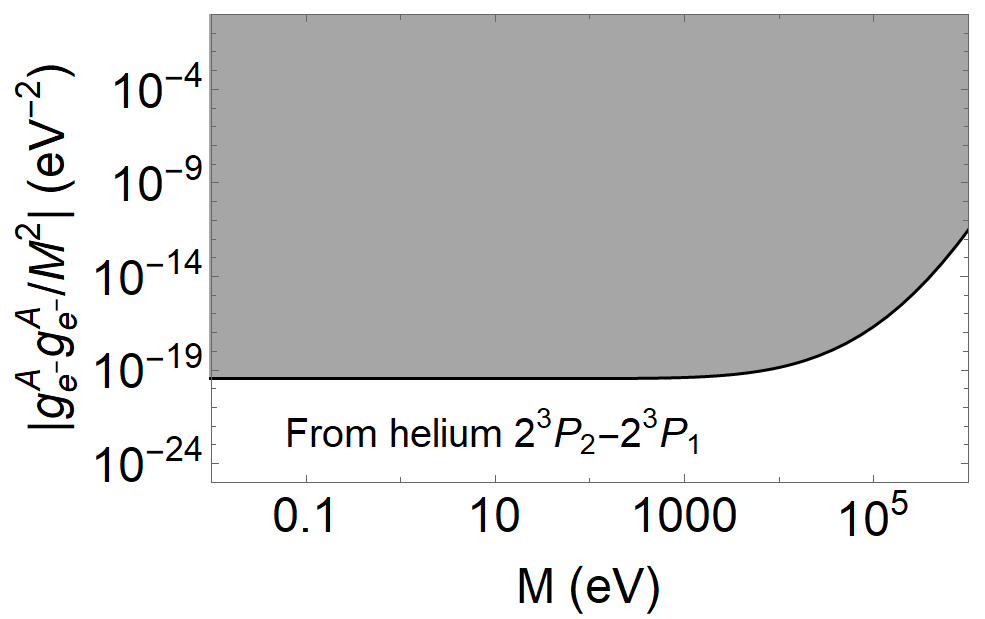}
\caption{Constraints for the interaction between electrons, at 90\% confidence level on the coupling constants as a function of boson mass, using $V_{AA}$ in Eq.\,\eqref{pseudovector-pseudovector_potential}.
The plot is based on experimental \cite{Marsman2015} and theoretical \cite{Pachucki2010} values for the $2^3P_2 - 2^3P_1$ transition in helium \cite{Ficek2017} and our numerical estimate of the spin-dependent contribution. 
} 
\label{fig:Helium}
\end{figure}

 \begin{figure} 
\includegraphics[width=0.46\textwidth]{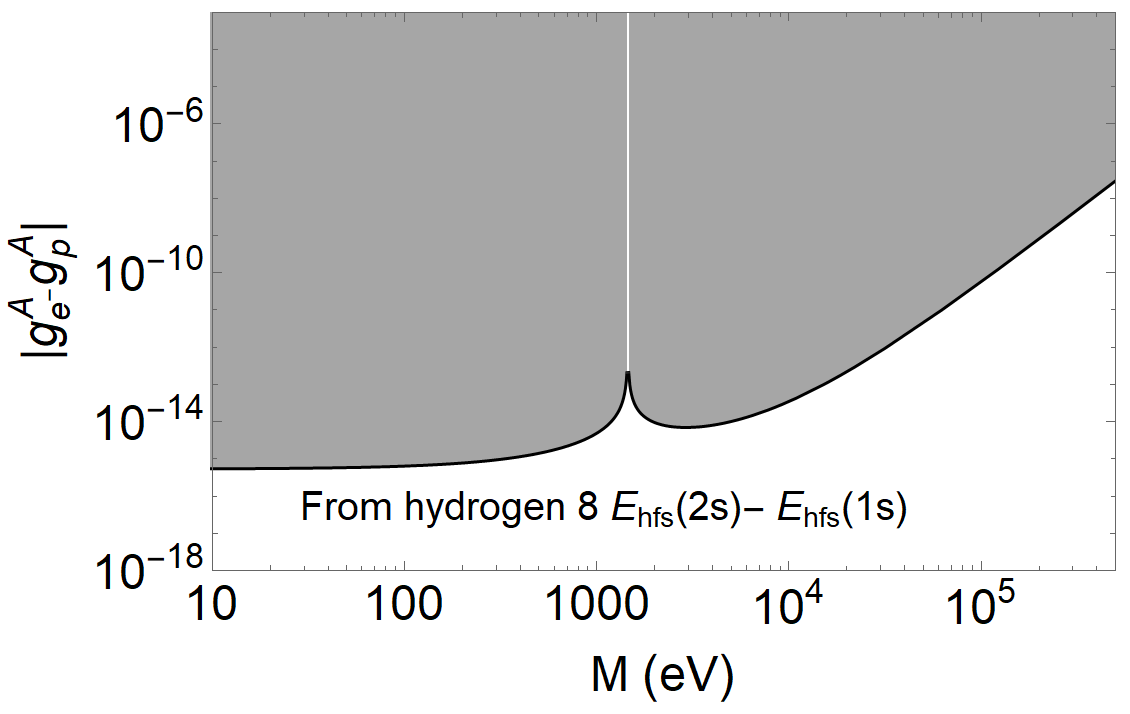}
\caption{Constraints for the electron-proton interaction, at 90\% confidence level on the coupling constants as a function of boson mass, using $V_{AA}$ in Eq.\,\eqref{pseudovector-pseudovector_potential}.
The plot is based on experimental 
\cite{Karshenboim2005,Hellwig1970,Zitzewitz1970,Essen1973,Morris71,Reinhard74,Petit80,Vanier76,Cheng80,Karshenboim00,Kolachevsky09}
and theoretical \cite{Ivanov} values for the $8 E_\mathrm{hfs} (2s)- E_\mathrm{hfs} (1s)$ difference between hyperfine transitions in hydrogen \cite{Karsh2011} and our numerical estimate of their spin-dependent contributions. 
Such a difference cancels the contribution of the contact terms, since the electron density on the proton in the $2s$ state is eight times smaller than in the $1s$ state.
The vertical asymptote at $1450$ eV is due to a cancellation in the denominator of Eq. \eqref{ggM2} for this plot.
} 
\label{fig:hydrogen}
\end{figure}

The solution we propose is to use a different form of the potential in numerical calculations, a form which appeared during the derivation of the potentials and contains the operator $\nabla$. Such a form for  Eq.\,\eqref{pseudoscalar-pseudoscalar_potential} is
\begin{equation}
\label{pseudoscalar-pseudoscalar_potential_form2}
V_{pp}(\v{r}) =  \frac{g_1^p g_2^p}{16 \pi m_1 m_2} \left( \v{\sigma}_1 \cdot \v{\nabla} \right) \left( \v{\sigma}_2 \cdot \v{\nabla} \right) \left( \frac{e^{-M r}}{r} \right)  \, .
\end{equation}
Then, calculating expectation values with Eq.\,\eqref{pseudoscalar-pseudoscalar_potential_form2}, we use integration by parts to avoid possible numerical issues of the contact term. From integration by parts of Eq.\,\eqref{pseudoscalar-pseudoscalar_potential_form2} we see that there is no physical problem, only a numerical one.

In Eq.\,\eqref{pseudoscalar-pseudoscalar_potential} the correct large-$M$ asymptotic 
is achieved due to delicate cancellation of different terms. This is hard to achieve in a numerical calculation. However, in Eq.\,\eqref{pseudoscalar-pseudoscalar_potential_form2} there is only one term, so no cancellation is required and the correct asymptotic is immediately seen 
$(e^{-Mr}/r \to \delta(\v{r}) \, 4 \pi/M^2)$. Using Eq.~\eqref{pseudoscalar-pseudoscalar_potential_form2} instead of Eq.~\eqref{pseudoscalar-pseudoscalar_potential} and integrating by parts, we arrive at Fig.\,\ref{fig:bound2} --- a bound on the $\mid g_{e^-}^p g_{\bar{p}}^p \mid$ coupling constants as a function of boson mass. Note that in \cite{Ficek2018} the bound was placed on the coefficient $f_3$, which relates to the pseudoscalar coupling constants in the following way \cite{Dob}:
$f_3 = -  \frac{g_{e}^p g_{\bar{p}}^p m_e}{4 m_{\bar{p}}}$ , where $m_e$ is the mass of the electron and $m_{\bar{p}}$ is the mass of the antiproton. 

We sort the potentials according to the type of mediating particle (scalar, vector, etc.)  and place limits on their coupling constants \cite{Fadeev2019}.
In this form the limits may be compared with the astrophysical, dark matter search and particle accelerator limits. 

\section{Results}
\subsection{New bound using 1/$\mathbf{\mathit{M}^2}$ term}
We use the properties discussed above to obtain a bound based on Eq.~\eqref{pseudovector-pseudovector_potential} for electron--antiproton interaction in antiprotonic helium. 
In order to avoid numerical issues as $M \to \infty$, the form of Eq.\,\eqref{pseudoscalar-pseudoscalar_potential_form2} can be used in calculating expectation values for the exclusion plot. Thus we construct Fig.\,\ref{fig:bound3}. To our knowledge, this is the first bound produced by the term proportional to $1/M^2$ in $V_{AA}$. 
Bounds on $V_{AA}$ of this type may be obtained  using the results in \cite{Ficek2017,Romalis,Romalisemail,Karshenboim,Karshenboim2010, Fru19}, or using any other scheme that is able to constrain $\mathcal{V}_3$.
Note further that the bound in Fig.~\ref{fig:bound3} is for a semileptonic spin-dependent interaction between matter (electron) and antimatter (antiproton).

\begin{table*}
\medskip \begin{tabular}{llll} \hline \hline
\rule{0ex}{3.0ex} Transition~~~~~~~~~~~~~~~~~~~~~~~~~~~~ & Bound ~~~~~~~~~~~~~~~~~~~~~~~~~~~~~~~~~~~~ & In the range ~~~~~~~~~~~ & In Figure \\
\hline
\rule{0ex}{2.8ex} Antiprotonic helium & $g_e^p g_{\bar{p}}^p \leq 0.0025$ & $M<10^2$~eV & \ref{fig:bound2} \\

\rule{0ex}{2.8ex} $ (35.5,35,34) - (34.5,34,34)$ & $g_e^A g_{\bar{p}}^A \leq 1.3 \times 10^{-18} (M /\text{eV})^2$ & $M<10^{2}$~eV & \ref{fig:bound3} \\
\hline

\rule{0ex}{2.8ex} Positronium & $g_{e^-}^p g_{e^+}^p  \leq 7.9 \times 10^{-6} $ & $M<10^2$~eV & \ref{fig:Positroniumvpvp} \\

\rule{0ex}{2.8ex} $1 ^3S_1 - 1 ^1S_0$ & $g_{e^-}^p g_{e^+}^p  \leq 1.0 \times 10^{-9} M/ \,\text{eV}$ & $M>10^5$~eV & \ref{fig:Positroniumvpvp} \\

\rule{0ex}{2.8ex}  & $g_{e^-}^A g_{e^+}^A \leq 7.5 \times 10^{-18}  (M /\text{eV})^2$ & $M<10^2$~eV & \ref{fig:Positronium} \\

\rule{0ex}{2.8ex}   & $g_{e^-}^A g_{e^+}^A \leq 2.5 \times 10^{-18}  (M /\text{eV})^2$ & $M>10^5$~eV &
\ref{fig:Positronium} \\
\hline
\rule{0ex}{2.8ex} Positronium & $g_{e^-}^p g_{e^+}^p \leq 7.2 \times 10^{-5}$ & $M<10^2$~eV & \ref{1s2sa} \\

\rule{0ex}{2.8ex} $1 ^3S_1 - 2 ^3S_1$ & $g_{e^-}^p g_{e^+}^p \leq 9.6 \times 10^{-9} M /\text{eV}$ & $M>10^5$~eV & \ref{1s2sa} \\

\rule{0ex}{2.8ex} & $g_{e^-}^A g_{e^+}^A \leq 6.9 \times 10^{-17}  (M /\text{eV})^2$ & $M<10^2$~eV & \ref{1s2sb} \\

\rule{0ex}{2.8ex}   & $g_{e^-}^A g_{e^+}^A \leq 2.3 \times 10^{-17}  (M /\text{eV})^2$ & $M>10^5$~eV & \ref{1s2sb} \\
\hline
\rule{0ex}{2.8ex} Muonium & $g_{e-}^p g_{\mu^+}^p  \leq 2.1 \times 10^{-8}$ & $M<10^2$~eV & \ref{fig:Muoniumvp} \\

\rule{0ex}{2.8ex} $1 ^3S_1 - 1 ^1S_0$ & $g_{e^-}^p g_{\mu^+}^p  \leq 1.4 \times 10^{-12} M /\text{eV}$& $M>10^5$~eV & \ref{fig:Muoniumvp} \\

\rule{0ex}{2.8ex}  & $g_{e^-}^A g_{\mu^+}^A \leq 9.5 \times 10^{-23}  (M /\text{eV})^2$ & $M<10^2$~eV & \ref{fig:Muonium} \\

\rule{0ex}{2.8ex} & $g_{e^-}^A g_{\mu^+}^A \leq 3.2 \times 10^{-23}  (M /\text{eV})^2$ & $M>10^5$~eV & \ref{fig:Muonium} \\
\hline
\rule{0ex}{2.8ex} Helium & $g_{e^-}^p g_{e^-}^p  \leq 4.4  \times 10^{-8}$ & $M<10^2$~eV & \ref{fig:Heliumpp} \\
\rule{0ex}{2.8ex}  $2^3P_2 - 2^3P_1$ & $g_{e^-}^A g_{e^-}^A  \leq 3.5 \times 10^{-20}  (M /\text{eV})^2$ & $M<10^2$~eV & \ref{fig:Helium} \\
\hline

\rule{0ex}{2.8ex} Hydrogen & $g_{e^{-}}^p g_{p}^p \leq 2.1  (M /\text{eV})^{-2} $ & $M<10^2$~eV & \ref{fig:hydrogenpp} \\

\rule{0ex}{2.8ex} $8 E_\mathrm{hfs} (2s) - E_\mathrm{hfs} (1s)$ & $g_{e^{-}}^p g_{p}^p \leq 1.8 \times 10^{-15} (M /\text{eV})^2$ & $M>10^5$~eV & \ref{fig:hydrogenpp} \\

\rule{0ex}{2.8ex} & $g_{e^{-}}^A g_{p}^A \leq 5.3 \times 10^{-16} $ & $M<10^2$~eV & \ref{fig:hydrogen} \\

\rule{0ex}{2.8ex}  & $g_{e^{-}}^A g_{p}^A \leq 4.5 \times 10^{-31}  (M /\text{eV})^4$ & $M>10^5$~eV & \ref{fig:hydrogen} \\
\hline \hline
\end{tabular}
\caption{Summary of the bounds obtained on properties of hypothetical bosons using various atomic systems.}
\label{Table:summary}
\end{table*}

The bound in Fig.~\ref{fig:bound3}, as well as bounds on figures below which use $V_{AA}$ are derived in the following way. The equivalent of Eq. \eqref{smallerthanE} for $V_{AA}$ is
\begin{equation} \label{gAgAbound}
    \Big| \, g_1^A g_2^A \left(\Delta U_2 (M) + \frac{1}{M^2} \Delta \tilde{U}_3 (M)\right) \, \Big| \leq \Delta E \, ,
\end{equation}
where $\Delta U_3 (M) = \Delta \tilde{U}_3 (M) / M^2$; $\Delta U_2 (M)$ and $\Delta U_3 (M)$ are related to $\mathcal{V}_2$ and $\mathcal{V}_3$ per Eq.\,\eqref{pseudovector-pseudovector_potential}. The bound in Fig.\,\ref{fig:bound3} is from
\begin{equation} \label{ggM2}
\Big| \, \frac{g_1^A g_2^A }{M^2} \, \Big| \leq \Big|  \frac{ \Delta E}{\left( M^2 \Delta U_2 (M) + \Delta \tilde{U}_3 (M)\right)} \Big|  \, .
\end{equation} 

The term $\Delta \tilde{U}_3 (M)$ dictates the shape of the plot for small mass $M$, while $ M^2 \Delta U_2 (M)$ dictates the shape for large mass $M$.
The ordinates differ between Fig.\,\ref{fig:bound2} and \,\ref{fig:bound3} since Eqs.\,\eqref{plotcreation} and \,\eqref{ggM2} are used, respectively. The scale of each figure is chosen to highlight the shape of each bound.

\subsection{Positronium, muonium, helium, and hydrogen}
We obtain a bound on the potential in Eq.\,\eqref{pseudovector-pseudovector_potential} using the ground-state $1 ^3S_1-1 ^1S_0$ transition in positronium. As in \cite{Derek2015}, we take $| \, \Delta E  \, | \leq 5 $ MHz \cite{Leslie2014}. The result appears in Fig.\,\ref{fig:Positronium} and its bound is described in Table \ref{Table:summary}. 
The shape of the bound line is explained by the fact that $\mathcal{V}_3$ dominates for small masses $M$, while $\mathcal{V}_2$ dominates for large masses $M$ where $ M^2 \Delta U_2 (M)$ results in a constant (see Appendix \ref{AnalyticalDerivation}).

We can get a bound on $g_{e^-}^A g_{e^+}^A$ from Eq. \eqref{gAgAbound}, instead of a bound on $g_{e^-}^A g_{e^+}^A / M^2$. Then we can compare the bound with the result in \cite{Derek2015} and see that we have a more stringent bound in the regime of $M\ll \Delta \tilde{U}_3 / \Delta U_2$.
This is due to the fact that, in contrast to Ref. \cite{Derek2015}, we use a potential containing the $1/M^2$ term.

In Figs.\,\ref{1s2sa} and \ref{1s2sb} we present bounds on pseudoscalar and pseudovector electron-positron interaction based on the $1^3 S_1 - 2^3 S_1$ transition in positronium.
We take $\Delta E = 10  \, \mbox{MHz}$ for this transition \cite{Fru19}. In Appendix \ref{AnalyticalDerivation} we give general analytical results for the potentials' expectation values in 1s and 2s states.
 
 The ground-state hyperfine transition is measured accurately also in the atomic system of muonium. Using this transition, we obtain a bound on the potential in Eq.~\eqref{pseudovector-pseudovector_potential}. As in \cite{Fru19}, we take $| \, \Delta E  \, | \leq 5 \times 10^{-4} $ MHz. The result appears in Fig.~\ref{fig:Muonium}. 

In Fig.\,\ref{fig:Helium} we obtain a bound on pseudovector coupling constants and boson mass from the $2^3P_2 - 2^3P_1$ transition of helium, using the results in \cite{Ficek2017}, where $| \, \Delta E  \, | \leq 3.7 $ kHz. 

Finally, in Fig.~\ref{fig:hydrogen} we use spectroscopic transitions in hydrogen to obtain a bound on electron-proton pseudovector interaction. Following \cite{Karsh2011} we take the difference (at 90\% confidence level) between theoretical and experimental results $| \, \Delta E  \, | \leq 0.102$ kHz for $8 E_\mathrm{hfs} (2s)- E_\mathrm{hfs} (1s)$, where $E_\mathrm{hfs}$ stands for the energy of the hyperfine transition in a particular state.
\\
\section{Conclusion}
One can search for new elementary particles using atomic spectroscopy. For the first time, we conduct such a search using a potential that results from the longitudinal polarization of a pseudovector particle. We also consider the pseudoscalar potential that includes a contact spin-dependent term, which does not contribute to new forces searched for in experiments with macroscopic objects, but does contribute in atomic spectroscopy.  
We extract limits on the interaction constants of pseudovector and pseudoscalar particles  from the experimental spectra of antiprotonic helium, muonium, positronium, helium, and hydrogen. The results are summarized in Table \ref{Table:summary}.

\textit{Acknowledgements} ---
We thank Derek Jackson Kimball, Claudia Frugiuele, Yevgeny V. Stadnik, Szymon Pustelny, Anne Fabricant, and Eric Adelberger for their valuable remarks. 
The authors acknowledge the support by the DFG Reinhart Koselleck project, the European Research Council Dark-OsT advanced grant under project ID 695405, and the Simons and the Heising-Simons Foundations. 
F. F. is supported by the Polish National Science Centre Grant No. 2020/36/T/ST2/00323.
V. V. F. is supported by the Australian Research Council  Grants No. DP190100974 and DP200100150 and the JGU Gutenberg Research Fellowship. 
M. G. K. is supported by RSF grant 19-12-00157 and is grateful to JGU for hospitality. 

 \appendix
 
 \section{Exclusion Plot with Contact Term} \label{ExclusionPlot}

\begin{figure}
\includegraphics[width=0.45\textwidth]{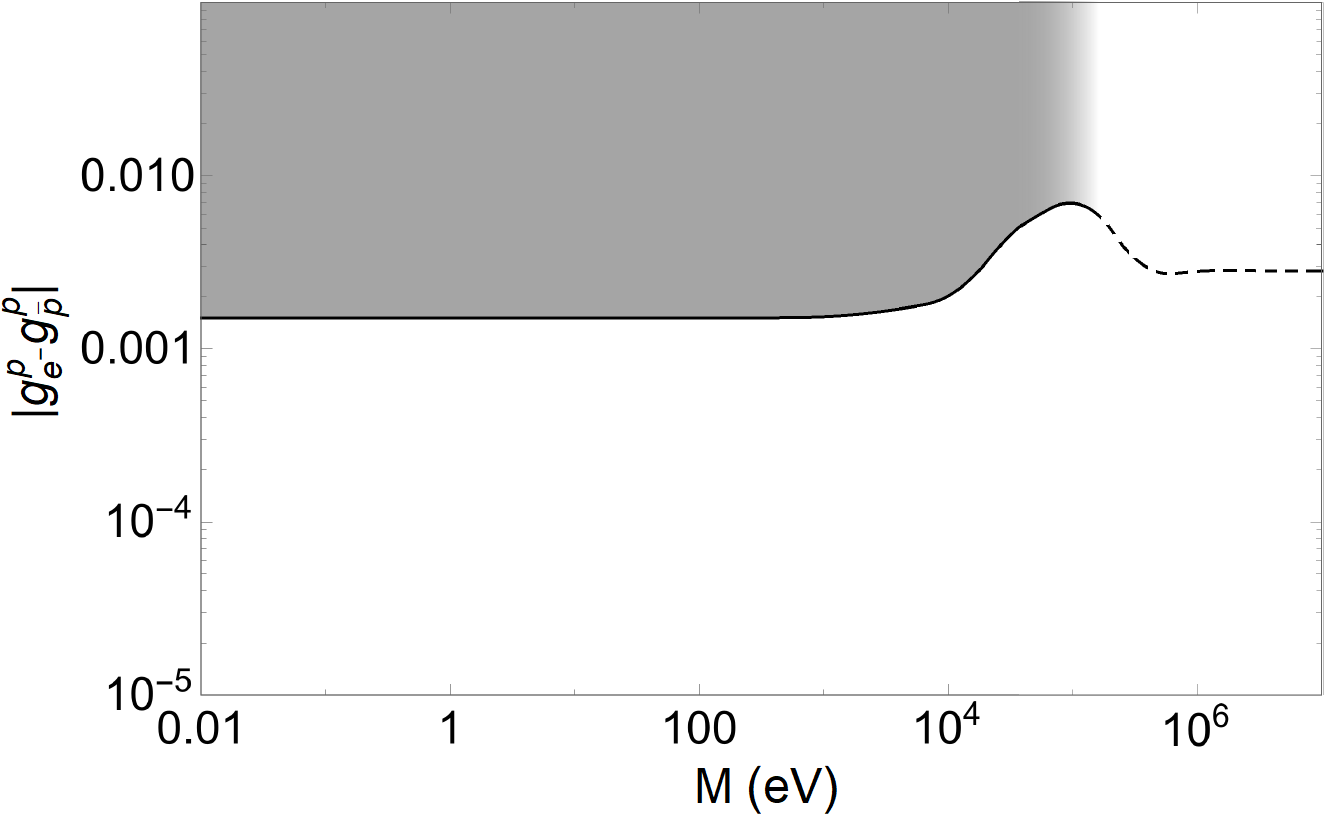}
\caption{Constraints for the interaction between an electron and an antiproton at 90\% confidence level on the coupling constants as a function of boson mass, using the $V_{pp}$ potential with the contact term [Eq.$(3)$ in the main text]
in numerical integration.
The bound for large masses $M$ is affected by 
absence of the proper cancellation between different terms in Eq.~$(3)$ of the main text.
The affected region on the top right is shown in white above a dashed line.
The shaded area is associated with the shaded area in Fig. 1(a) in the main text. See Fig. 1(a) for the accurate bound.
} 
\label{fig:bound1}
\end{figure}

Direct application of Eq. (3) in the main text leads to Fig.~\ref{fig:bound1}, where apparently we obtained a bound on the coupling constants for any boson mass $M$, as the bound edge is horizontal on the right side of the plot. Nonetheless, this bound plot is incorrect for boson masses much larger than the fermion masses, due to numerical reasons.
The problem is that the calculation for large masses $M$ is affected by 
absence of the proper cancellation between different terms in Eq. $(3)$ of the main text.
Therefore in Fig.~\ref{fig:bound1} we colored in white the bound where the result is inaccurate. 

By focusing on $M<m_1,m_2$ (where $m_1$ and $m_2$ are fermion masses) we avoided the issue of finite numerical precision at large boson masses in the exclusion plot of Fig. 3 (b) in \cite{Ficek2018}. This ensured that the plot in \cite{Ficek2018}, which includes the contribution of the contact term, is correct in the range considered.

\section{Analytical Derivation of Expectation Values}\label{AnalyticalDerivation}
Consider the potentials without their coupling constants coefficients
 \begin{widetext}
\begin{equation*}
    V_2=(\mathbf{\sigma}_1\cdot\mathbf{\sigma}_2)  \frac{e^{-Mr}}{r},\qquad
    V_3=\left[\mathbf{\sigma}_1\cdot\mathbf{\sigma}_2\left(\frac{M}{ r^2}+\frac{1}{r^3}+\frac{4\pi}{3}\delta^3(\mathbf{r})\right)-(\mathbf{\sigma}_1\cdot\mathbf{\hat{r}})(\mathbf{\sigma}_2\cdot\mathbf{\hat{r}})\left(\frac{M^2}{ r}+\frac{3M}{ r^2}+\frac{3}{r^3}\right)\right]e^{-Mr} \, .
\end{equation*}
 \end{widetext}
We need the impact of these potentials on the energy difference between the $1^3 S_1$ and $2^3 S_1$ states in hydrogen, muonium and positronium, which are spherically symmetric. This allows us to average the $V_3$ potential over angles, using 
$\langle \hat{r}_i \hat{r}_k \rangle = \frac13 \delta_{ik}$. 
Note also that $\langle \mathbf{\sigma}_1\cdot\mathbf{\sigma}_2 \rangle=1$ for the total spin $S=1$ states. As a result we only need integration of the potentials 
\begin{equation*}
    \langle V_2 \rangle= \frac{e^{-Mr}}{r},\qquad \langle V_3 \rangle=\frac{1}{3} \left( 4 \pi \delta(r) - \frac{M^2}{r} \right)e^{-Mr} \, ,
\end{equation*}
with the squared hydrogen-like wave functions for $1s$ and $2s$ orbitals 
\begin{align}
    |\psi_1(r)|^2 =\frac{k^3e^{-2kr}}{\pi},\qquad |\psi_2(r)|^2=\frac{k^3 e^{-kr}}{ 8 \pi}  \left(1-\frac{kr}{2}\right)^2,
\end{align}
where $k=1/a$ for hydrogen and muonium and $k=1/2a$ for positronium, where $a$ is the Bohr radius. For hydrogen-like ions $k=Z/a$. The results are 
\begin{align} \label{expectvaluesV2}
    \langle\psi_1|V_2|\psi_1\rangle &= \frac{4 k^3}{(2k +M)^2} \, , \nonumber \\
    \langle\psi_2|V_2|\psi_2\rangle &= \frac{ k^3(k^2+2 M^2)}{4( k +M)^4} \, ,
\end{align}
\begin{align} \label{expectvaluesV3}
    \langle\psi_1|V_3|\psi_1\rangle &= \frac{16 k^4(k+M)}{3(2 k +M)^2} \, , \nonumber \\
    \langle\psi_2|V_3|\psi_2\rangle &= \frac {k^3}{6} - \frac{ k^3 M^2(k^2+ 2 M^2)}{12(k +M)^4} \, .
\end{align}

\section{Additional plots of bounds on pseudoscalar interactions}
\begin{figure} [h!]
\includegraphics[width=0.46\textwidth]{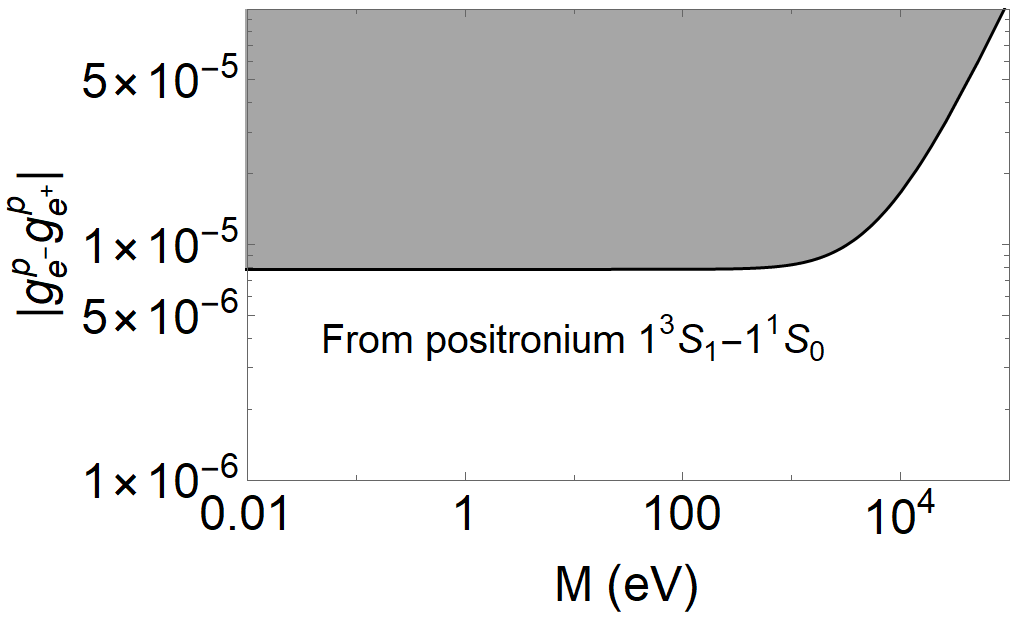}
\caption{Constraints for the electron-positron interaction at 90\% confidence level on the coupling constants as a function of boson mass using Eq.(3) in the main text.
The plot is based on experimental \cite{Ishida2014,Mills1975,Ritter1984} and theoretical \cite{Eides2014} values for the $1 ^3S_1-1 ^1S_0$ ground state transition in positronium \cite{Derek2015} and our numerical estimate of the spin-dependent contribution. 
} 
\label{fig:Positroniumvpvp}
\end{figure}

\begin{figure} [h!]
\includegraphics[width=0.46\textwidth]{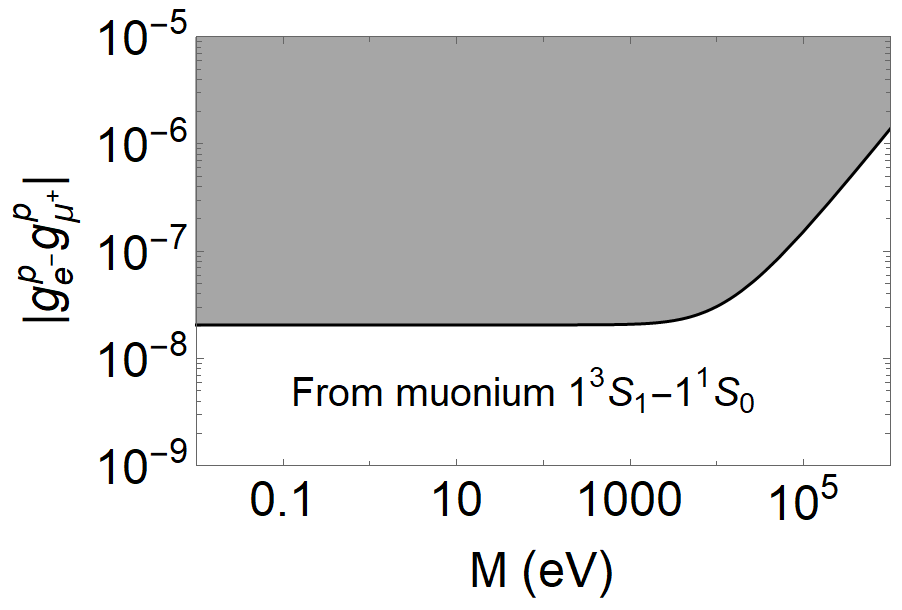}
\caption{Constraints for the interaction between an antimuon and an electron, at 90\% confidence level on the coupling constants as a function of boson mass, using $V_{pp}$ in Eq. (3) of the main text.
The plot is based on experimental \cite{Liu1999} and theoretical \cite{Karshenboim2014,Karshenboim2005} values for the hyperfine ground state transition in muonium \cite{Fru19} and our numerical estimate of the spin-dependent contribution. 
} 
\label{fig:Muoniumvp}
\end{figure}

\begin{figure} [h!]
\includegraphics[width=0.46\textwidth]{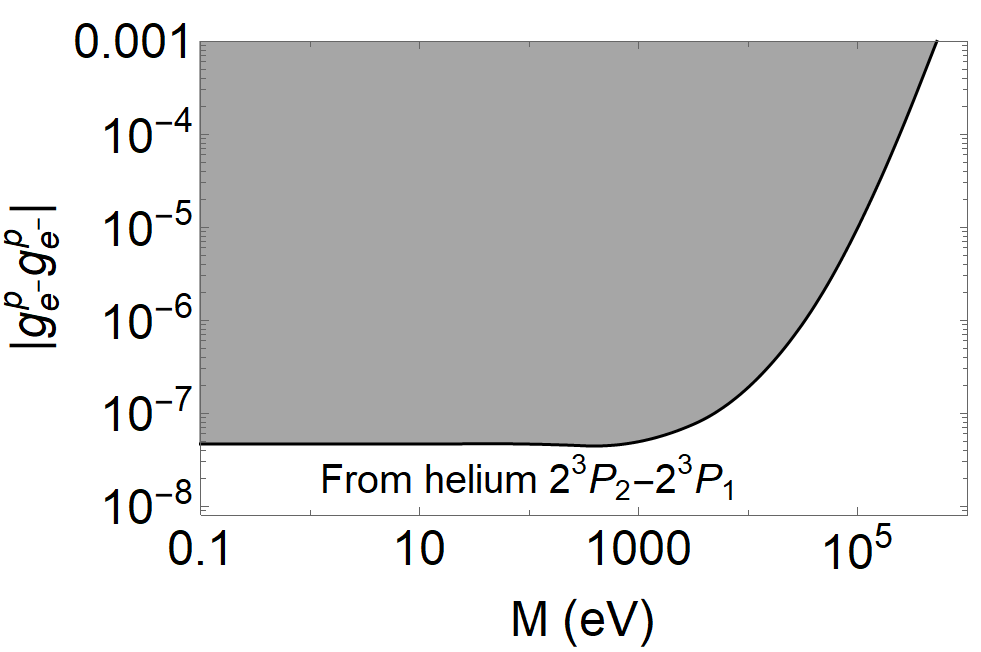}
\caption{Constraints for the interaction between electrons, at 90\% confidence level on the coupling constants as a function of boson mass, using $V_{pp}$ in Eq.\,$(3)$ of the main text.
The plot is based on experimental \cite{Marsman2015} and theoretical \cite{Pachucki2010} values for the $2^3P_2 - 2^3P_1$ transition in helium \cite{Ficek2017} and our numerical estimate of the spin-dependent contribution. 
} 
\label{fig:Heliumpp}
\end{figure}

\begin{figure} [H]
 \centering
\includegraphics[width=0.46\textwidth]{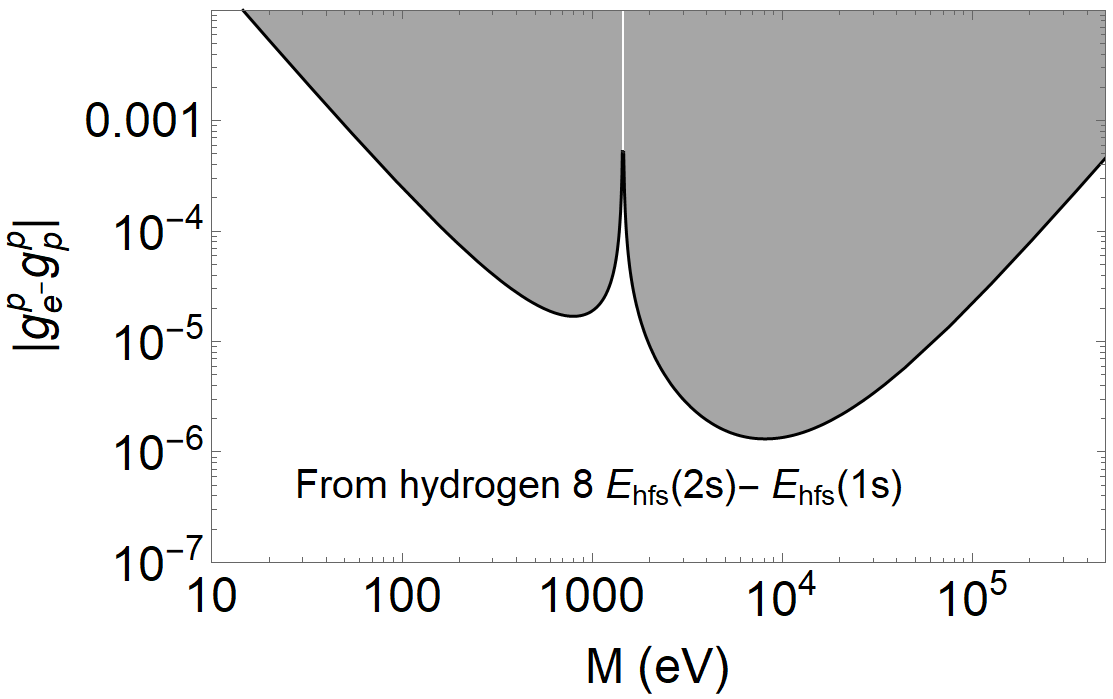}
\caption{Constraints for the electron-proton interaction, at 90\% confidence level on the coupling constants as a function of boson mass, using $V_{pp}$ in Eq.(3) of the main text.
The plot is based on experimental \cite{Karshenboim2005,Hellwig1970} and theoretical \cite{Ivanov} values for the $8 E_\mathrm{hfs} (2s)- E_\mathrm{hfs} (1s)$ difference between hyperfine transitions in hydrogen \cite{Karsh2011} and our numerical estimate of their spin-dependent contributions.  
} 
\label{fig:hydrogenpp}
\end{figure}

In Figures
\ref{fig:Positroniumvpvp},
\ref{fig:Muoniumvp},
\ref{fig:Heliumpp}, and \ref{fig:hydrogenpp}
we show several plots referred to in Table I of the main text.

 \


\end{document}